\documentclass[floatfix,aps,prl,superscriptaddress, twocolumn]{revtex4-2}
\usepackage{braket}
\usepackage{amsmath}
\usepackage{amssymb}
\usepackage{svg}
\usepackage[utf8]{inputenc}
\usepackage[T1]{fontenc}
\usepackage{lineno}
\usepackage{comment}
\usepackage{graphicx}
\usepackage[normalem]{ulem}
\usepackage[colorlinks,breaklinks]{hyperref}
\usepackage{siunitx}
\usepackage{soul}
\definecolor{darkred}{rgb}{0.5,0,0}
\definecolor{darkgreen}{rgb}{0,0.5,0}
\definecolor{darkblue}{rgb}{0,0,0.5}
\hypersetup{linkcolor=darkblue,citecolor=darkblue,urlcolor=darkblue}

\usepackage[toc,page]{appendix}
\usepackage{tikz}
\usetikzlibrary{math}
\usetikzlibrary{decorations.pathreplacing,calligraphy}

\begin{document}
% \nolinenumbers
\title{Observing High-Dimensional Bell Inequality Violations using Multi-Outcome Spectral Measurements}

\author{Kiki Dekkers}
\affiliation{School of Engineering and Physical Sciences, Heriot-Watt University, Edinburgh, EH14 4AS, UK}

\author{Laura Serino}
\affiliation{Paderborn University, Integrated Quantum Optics, Institute for Photonic Quantum Systems (PhoQS), Paderborn, Germany}

\author{Nicola D'Alessandro}
\affiliation{Physics Department and NanoLund, Lund University, Box 118, 22100 Lund, Sweden\\ $^*$j.leach@hw.ac.uk}

\author{Abhinandan Bhattacharjee}
\affiliation{Paderborn University, Integrated Quantum Optics, Institute for Photonic Quantum Systems (PhoQS), Paderborn, Germany}

\author{Benjamin Brecht}
\affiliation{Paderborn University, Integrated Quantum Optics, Institute for Photonic Quantum Systems (PhoQS), Paderborn, Germany}

\author{Armin Tavakoli}
\affiliation{Physics Department and NanoLund, Lund University, Box 118, 22100 Lund, Sweden\\ $^*$j.leach@hw.ac.uk}

\author{Christine Silberhorn}
\affiliation{Paderborn University, Integrated Quantum Optics, Institute for Photonic Quantum Systems (PhoQS), Paderborn, Germany}

\author{Jonathan Leach}
%\email{j.leach@hw.ac.uk}
\affiliation{School of Engineering and Physical Sciences, Heriot-Watt University, Edinburgh, EH14 4AS, UK}

%\linenumbers

\begin{abstract}

Violation of Bell inequalities is an essential requirement for many quantum information and communication protocols. In high-dimensional systems, Bell inequality tests face the challenge of implementing genuinely multi-outcome measurements, since the emulation of these with separate dichotomic projections opens a binarisation loophole that local hidden variable theories can exploit. Here we show that the joint spectral intensity of a two-photon entangled state contains access to the necessary multi-outcome measurements to overcome this obstacle and violate a Bell inequality for high-dimensional states. This result is contrary to the belief that the joint spectral intensity is a phase-insensitive quantity and does not have sufficient information to certify entanglement or Bell-nonlocality. Using this approach, we violate the CGLMP Bell inequality up to dimension $d$ = 8, all with negligible $p$-values, and for the first time close the binarisation loophole in high-dimensional Bell experiments. Guaranteeing Bell-nonlocal correlations using frequency-only measurements removes the technological hurdle of measurements in the temporal domain, thus greatly simplifying any practical implementation of future high-dimensional quantum information protocols.

\end{abstract}

\maketitle

\begin{figure*}
 \centering \includegraphics{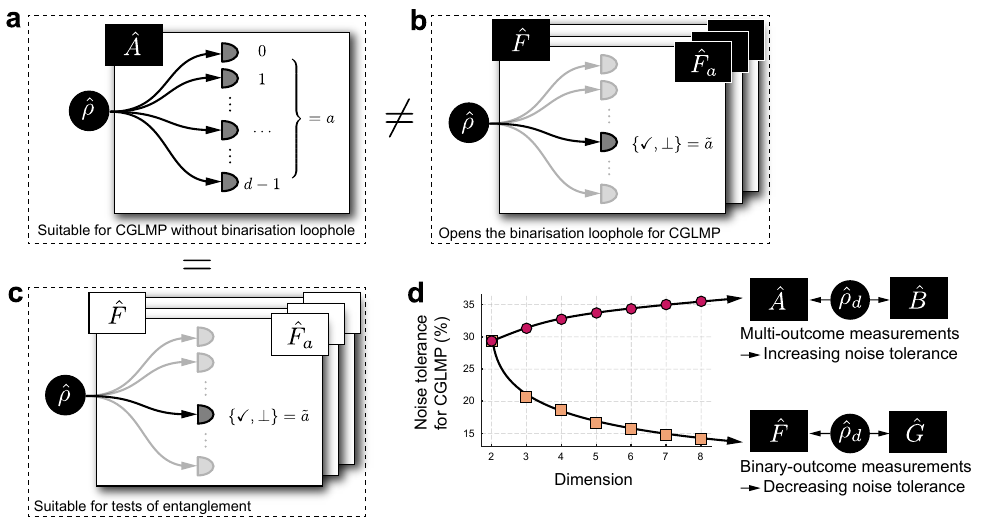} 
 \caption{\textbf{Multi-outcome vs binarised measurements and implications for noise tolerance of high-dimensional Bell tests.} \textbf{a}, Multi-outcome measurement. The measurement resolves all $d$ possible outcomes in every round. \textbf{b}, Binarised emulation of multi-outcome measurement $\hat{A}$. A sequence of measurements, $\hat{F}_a$, is implemented with each outcome corresponding to whether detector number $a$ clicks or not. This is the standard approach for high-dimensional Bell tests but opens up the binarisation loophole for uncharacterised measurements (black box). \textbf{c}, If the hypothesis being tested permits characterised measurements (white box), e.g.~entanglement witness tests, the measurement $\hat{A}$ can be implemented by the procedure $\hat{F}$. \textbf{d}, Noise tolerance levels for the CGLMP test for multi-outcome vs.~binarised measurements.  The noise tolerance increases with dimension only for the multi-outcome scenario.}
 \label{fig:binary_vs_multi}
\end{figure*}

The concept that a quantum state cannot be defined with zero uncertainty in more than one basis is the cornerstone of the uncertainty principle and fundamental to the quantum correlations that exist for entangled states. This also plays a central role in the violation of Bell inequalities that are essential for the performance of many quantum information protocols.  Many emerging quantum technologies rely on self-testing to guarantee their task performance. For example, for performing quantum key distribution (QKD) using untrusted devices, the state shared between Alice and Bob must be able to demonstrate Bell-nonlocality, i.e. the absence of local causality \cite{acin_bells_2006}. %be nonlocally correlated. 
The violation of a Bell inequality \cite{bell_einstein_1964} rules out local hidden variable models and, thus, guarantees such Bell-nonlocal correlations, and consequently entanglement. In high-dimensional systems, which are a major frontier in quantum information due to increased information capacities and noise resistance, the hallmark test for Bell-nonlocality is the Collins-Gisin-Linden-Massar-Popescu (CGLMP) Bell inequality~\cite{collins_bell_2002}.

High-dimensional Bell inequality violations have been realised in photonics using entanglement in time-bins \cite{thew_bell-type_2004, ikuta_enhanced_2016}, orbital angular momentum (OAM) \cite{dada_experimental_2011}, number of entangled pairs \cite{lo_experimental_2016}, and path \cite{wang_multidimensional_2018, hu_observation_2025}. For certain quantum information tasks, high-dimensional entangled states, which exist in a $d$-dimensional Hilbert space, offer advantages for information encoding \cite{Hu_2018}, reduced efficiency thresholds for the detection loophole \cite{Vertesi_2010, hu_high-dimensional_2022}, enhanced robustness to noise \cite{kaszlikowski_violations_2000, zhu_is_2021, srivastav_quick_2022}, and improved security in QKD protocols \cite{bechmann-pasquinucci_quantum_2000, cerf_security_2002, durt_security_2004, sheridan_security_2010}. 
Many quantum information tasks now utilise high-dimensional entanglement \cite{Bruss_2002, wang_quantum_2005, zhong_photon-efficient_2015, sit_high-dimensional_2017, Luo_2019} and methods have been developed to controllably generate \cite{Morrison2022, serino_orchestrating_2024} and efficiently detect this \cite{Bavaresco_2018, Morelli2023, Cobucci2024, chang_2024}. QKD was recently reported with high-dimensional time-bin entangled photons \cite{yu_quantum_2025}.

In a high-dimensional Bell test, Alice and Bob privately and independently select measurements $x$ and $y$ and record the outcomes $a$ and $b$ respectively, each taking one of $d$ possible values. The basis vectors of Alice and Bob's measurements $\{\hat{A}_{a|x}\}$ and $\{\hat{B}_{b|y}\}$ are always chosen to reveal any difference between the predictions of quantum theory and any local hidden variable theory, and the experiment is repeated many times with the aim of proving that the emerging joint probability distribution $P(a,b|\hat{A},\hat{B})$ does not admit a local hidden variable model.
Once the measurements have been made, the observed correlations are compared to the maximum correlation strength allowed by local hidden variables through the appropriate Bell inequality. 

As high-dimensional multi-outcome measurements are experimentally challenging, it has been standard practice to use a binarised implementation of measurements approach to measurements for high-dimensional Bell violations, see for example references \cite{dada_experimental_2011, wang_multidimensional_2018, hu_observation_2025}. In such a binarised implementation, the multi-outcome measurements $\hat{A}$ and $\hat{B}$ are substituted with a sequence of measurements with only binary outcomes, $\hat{F}$ and $\hat{G}$. Thus, instead of projecting the system randomly onto one of the $d$ outcomes (Fig.~\ref{fig:binary_vs_multi}a), the procedure $\hat{F}$ ($\hat{G}$) projects the system onto each of the possible outcomes via $d$ separate measurements (Fig.~\ref{fig:binary_vs_multi}b).  A successful projection is labelled $\tilde{a}=\mkern-4mu\checkmark$ ($\tilde{b}=\mkern-4mu\checkmark$) and a failed one  (``no-click'') is labelled $\tilde{a}=\perp$ ($\tilde{b}=\perp$).  However, it was recently realised that substituting multi-outcome measurements for binarised measurements in high-dimensional Bell experiments, such as the CGLMP inequality test, leads to a loophole in the falsification of local hidden variable models \cite{tavakoli_binarization_2025}.  

The reason for the loophole is that a binarised measurement implementation assumes that $P(\tilde{a},\tilde{b}|\hat{F},\hat{G}) = P(a,b|\hat{A},\hat{B})$, but this cannot be guaranteed for the uncharacterised measurement devices that are required in Bell tests.   Such an assumption is justified only when the theoretical models that we aim to falsify, e.g.~separability, can be tested with pre-determined measurements, see Fig.~\ref{fig:binary_vs_multi}c.   In order to not artificially constrain local hidden variable theories, high-dimensional Bell tests require that each binary-outcome measurement must be treated as an independent uncharacterised setting.  The binarised approach, which generates $P(\tilde{a},\tilde{b}|\hat{F},\hat{G})$, requires Alice and Bob to not only make a choice of which basis to measure but also which outcome to examine, i.e.~the number of choices that Alice and Bob each make has increased by a factor $d$. Since the two distributions $P(a,b|\hat{A},\hat{B})$ and $P(\tilde{a},\tilde{b}|\hat{F},\hat{G})$  have different numbers of inputs and outputs, they belong to distinct different correlation spaces, and their Bell-nonlocality must therefore be detected via qualitatively different Bell inequalities. 

One can still find Bell-nonlocality in the distribution $P(\tilde{a},\tilde{b}|\hat{F},\hat{G})$, but this is found to be weaker and less tolerant to noise than that found in $P(a,b|\hat{A},\hat{B})$; see Fig.~\ref{fig:binary_vs_multi}d. As an example, when certifying Bell-nonlocality in $d$ = 8 using the CGLMP inequality using an optimally entangled state \cite{Zohren_2008}, multi-outcome measurements can tolerate up to 35.5\% noise, whereas binary-outcome measurements can only tolerate 14.9\% noise \cite{tavakoli_binarization_2025}. The gains of increased noise tolerance of high-dimensional states are no longer realised, and, thus, it is essential that the binarisation loophole is closed in real-world, noisy environments.

\begin{figure*}
 \centering \includegraphics{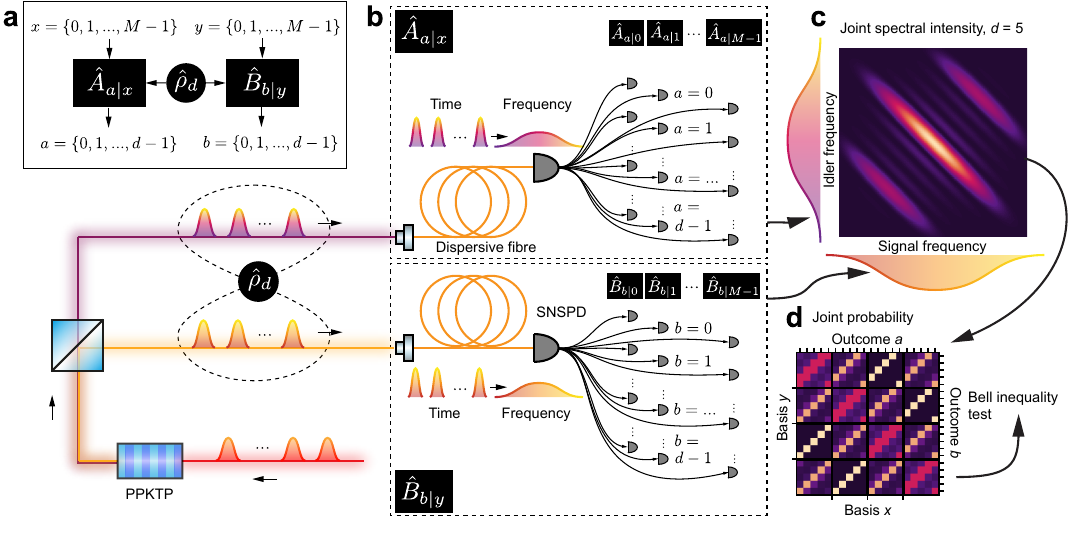}
 \caption{\textbf{Schematic of high-dimensional Bell tests and experimental implementation using multi-outcome spectral measurements.} \textbf{a}, The state $\hat{\rho}_d$ is directed to untrusted multi-outcome measurement devices $\hat{A}_{a|x}$ and $\hat{B}_{b|y}$ to establish the joint probabilities $P(a,b|x,y)$. \textbf{b}, In the experiment, high-dimensional states are generated by SPDC using a temporally shaped pulse passed through a periodically poled potassium titanyl phosphate (PPKTP) waveguide. The signal and idler photons are separated with a polarising beam splitter (PBS) and passed through dispersive fibre such that the arrival time at the superconducting nanowire single-photon detectors (SNSPD) corresponds to a measurement in the frequency domain. \textbf{c}, The joint spectral intensity of the signal and idler is measured.  For two-photon high-dimensional input states, this has interference fringes.  \textbf{d}, The JSI is converted to the joint probabilities $P(a,b|x,y)$, which can be applied to Bell inequality tests.}
 \label{fig: Schematic}
\end{figure*}

In the context of temporal entanglement, implementing the multi-outcome measurements in multiple bases required by the CGLMP inequality becomes increasingly challenging in high dimensions. While direct measurements of time-bins in the nanosecond regime can be achieved through simple arrival time detection, projections onto superposition bases require complex phase-sensitive techniques. These include cascaded interferometers \cite{islam_robust_17, vagniluca_efficient_20}, nonlinear frequency conversion processes \cite{tang_multiplexed_24, serino_realization_23, serino_programmable_25}, or combinations of electro-optic modulators and pulse shapers \cite{lu_electrooptic_18, lu_frequency_23}. Measuring the frequency of single-photon states is straightforward using spectrometers with photon counting detectors. Time-of-flight spectrometers are a resource-efficient way of implementing such spectral measurements \cite{avenhaus_fiber_09, widomski_efficient_24}.

Traditionally, these joint spectral intensity measurements have been regarded as insufficient to verify nonlocality or entanglement because they are not phase sensitive and were therefore thought to probe only a single basis. Measurements in two bases \cite{Bavaresco_2018}, time and frequency in this case, would be required to establish entanglement or quantum steering \cite{chang_2024}. However, for discrete time-bins the joint spectral intensity contains interference fringes, and in this case, not only does it contain sufficient information to establish entanglement, but it also contains the necessary information to rule out local hidden variables and establish Bell-nonlocality. The measurement can be understood as projections onto different time-bin superposition states, and a simple time-of-flight spectrometer is one example of a system capable of realizing time-bin superposition measurements, provided that the fibre is long enough to reach the far field \cite{avenhaus_fiber_09, widomski_efficient_24}.
 
In this work, we address both the binarisation loophole and reveal that the joint spectral intensity of frequency-only measurements contains sufficient information to rule out local hidden variable models via the violation of the CGLMP inequality.  We demonstrate the first high-dimensional Bell inequality violation using multi-outcome measurements for two-photon temporal mode states with controllable dimensionality \cite{serino_orchestrating_2024}. 
We show all necessary measurement bases can be accessed in the frequency domain, thus removing the need for any measurements in the temporal degree of freedom. This represents a significant advance over the standard approach, where high-dimensional Bell tests have been conducted using binary-outcome measurements, where only detection events for one of the $d$ possible outcomes in a given measurement basis are recorded at a time. Moreover, frequency measurements alone are sufficient to demonstrate entanglement, and this drastically simplifies any implementation of any quantum information protocol, such as QKD, that uses correlations in the time/frequency domain.
\begin{figure*}
 \centering
 \includegraphics{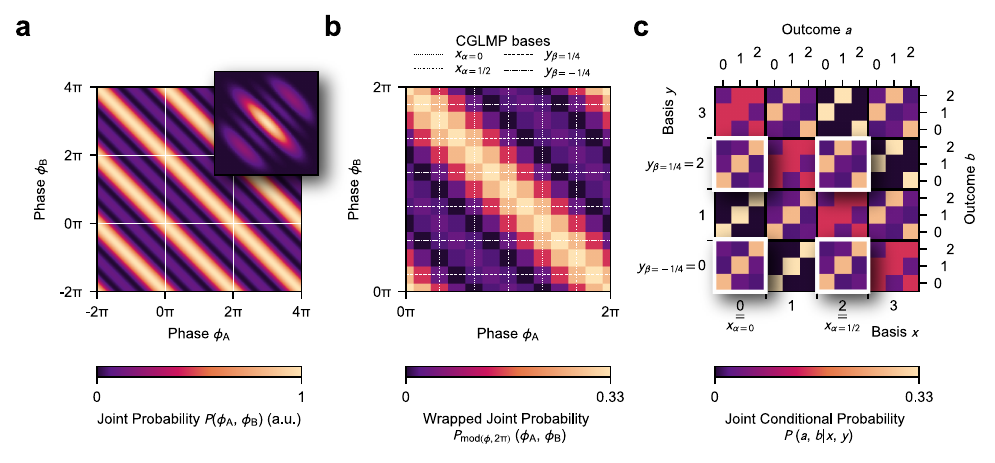}
 \caption{\textbf{Theoretical joint probability distributions of a 3-dimensional maximally entangled state. }\textbf{a}, Continuous joint probability in its Fourier space. The white lines indicate periodic $\phi$ regions with period 2$\pi$. The inset is the same distribution but representative of an experimental measurement.  The finite width of the basis states results in an envelope that modulates the joint probability. \textbf{b}, Wrapped joint probability distribution.  The intersections of the white lines indicate the measurements required for the CGLMP inequality. The resolution in this example is chosen to be low to highlight the measurement procedure. The number of outcomes in a $2 \pi$ window is $N$ = 12, so that the number of bases is $M$ = 4.      \textbf{c}, The wrapped joint probability can be mapped to the conditional probabilities $P(a,b|x,y)$, and the white squares indicate the probabilities used for the CGLMP inequality. }
 \label{fig: Theoretical Simulations}
\end{figure*}
\section{Results}

\subsection{Summary of Methods}

We define a set of discrete temporal modes for Alice and Bob to establish $d$-dimensional entanglement \cite{serino_orchestrating_2024}. This is then mapped onto its Fourier space via the use of dispersive fibres. The Fourier transform of a discrete set of basis vectors is periodic, parameterised by the phase~$\phi$. This means that a measurement in the Fourier transform space of the computational basis reveals all equally weighted real and imaginary superpositions of the computational basis vectors. This is the core principle of our system, as shown in Fig.~\ref{fig: Schematic}.  The obtained joint spectral intensity thus contains a set of measurement bases and outcomes for Alice and Bob that contain all the necessary measurements required to test the measured correlations against local hidden variable models.

\subsection{Theory}

Consider a $d$-dimensional maximally entangled state of the form
\begin{equation}
\ket{\psi_d} = \frac{1}{\sqrt{d}} \sum_{j=0}^{d-1} \ket{j}_\mathrm{A} \ket{j}_\mathrm{B}, 
\label{eq:max_entangled_state}
\end{equation}
where $\ket{j}_\mathrm{A}$ are Alice's basis states, $\ket{j}_\mathrm{B}$ are Bob's basis states. As the state space of Alice and Bob's is discrete, the Fourier space is continuous and periodic. If Alice and Bob both perform Fourier transforms, the joint probability $P(\phi_\mathrm{A}, \phi_\mathrm{B}) = |\bra{\phi_\mathrm{A}}\bra{\phi_\mathrm{B}} \psi \rangle|^2$ is given by
\begin{align}
P(\phi_\mathrm{A}, \phi_\mathrm{B}) = \frac{1}{d^3} \frac{\sin^2 \left[d(\phi_\mathrm{A}+\phi_\mathrm{B})/ 2 \right]}{\sin^2 \left[ (\phi_\mathrm{A}+\phi_\mathrm{B}) / 2 \right]},
\label{eq:theoretical_JPD_wrapped}
\end{align}
where $\ket{\phi} = \frac{1}{\sqrt{d}} \sum_{j=0}^{d-1} \exp\left(- i j \phi \right) \ket{j}$.
Fig.~\ref{fig: Theoretical Simulations}a shows the joint probability for $d$ = 3 for Alice and Bob. We see that the joint probability has a periodicity of 2$\pi$ and is maximised for values of $\phi_\mathrm{A}+\phi_\mathrm{B} = n \times 2 \pi$, where $n$ is an integer.

This concept is related to the theoretical work of Li and Zhao \cite{li_bells_2018} who, in the context of position/momentum entanglement, highlighted that a Bell inequality can be violated for a system entangled in a Hilbert space by only conducting measurements in its Fourier transform-related dual space. Analogous to Young's double slit experiment, detection in the far-field of a set of $d$ slits corresponds to the superposed projective measurement of the $d$ slit states, each with a relative phase difference. 

In an experimental test, our goal is to establish $P(a,b|x,y)$  from the coincidence measurements $C(\phi_\mathrm{A}, \phi_\mathrm{B})$, where $x$ is the choice of basis for Alice, and $y$ is the choice of basis for Bob.  As the basis choices, $x$ and $y$ are made passively, the experiment first measures $P(a,b,x,y)$ from which we establish the conditional probability $P(a,b|x,y)$. This is a manifestation of the detection loophole and is similar to the work of Gisin \cite{tittel_violation_1998}, who violated a Bell inequality using passive basis choice.  As noted in that work, there could exist a hidden variable that determines which basis to collapse into, and this stands in sharp contrast to the random choice described in quantum mechanics. Alice or Bob could make active and independent choices of their measurement basis. This would necessitate disregarding 50\% of their outcomes, but, assuming the fair sampling loophole, the joint distribution $P(a,b|x,y)$ is valid.  A full analysis of active vs.~passive basis choice and its implications on the locality and detection loopholes can be found in ref.~\cite{gisin_bell_1999}.

Although $\phi$ is a continuous parameter, in practice, finite resolution for Alice and Bob's detectors will discretise the measurements. If, in any $2\pi$ window, the experiment records $N$ outcomes, these can be identified as the $d$ outcomes from $M$ measurement bases, i.e.~$N = M\times d$. This means that rather than project onto the continuous $\ket{\phi}$ states, the experiment outcomes correspond to an integration over the range $\phi$ to $\phi+2 \pi/N$.  The discretised space of $\phi_\mathrm{A}$ and $\phi_\mathrm{B}$ is shown in Fig.~\ref{fig: Theoretical Simulations}b, where $N$~=~12, so that $M$ = 4. If Alice and Bob record coincidences over multiple $2\pi$ regions $C(\phi_\mathrm{A}, \phi_\mathrm{A})$, these outcomes can be wrapped back to a single $2\pi$ region $C_{
\mathrm{mod}(\phi,\, 2\pi)}(\phi_\mathrm{A}, \phi_\mathrm{A})$, and mapped to a conditional probability $P(a, b|x,y)$, see supplementary information for further details. Importantly, all the measurement outcomes and bases required for testing the data against local hidden variables are contained within the wrapped version of the joint probability $P_{
\mathrm{mod}(\phi,\, 2\pi)}(\phi_\mathrm{A}, \phi_\mathrm{B})$ and therefore $P(a,b|x,y)$.

The mapping that is necessary for the CGLMP inequality is given by 
\begin{align}
\phi_{A} & \rightarrow \frac{2\pi}{d}\left(a + \frac{x}{M}\right) {\rm~and}~\\
\phi_{B} & \rightarrow \frac{2\pi}{d}\left(\mathrm{mod}(-b,d) + \frac{y}{M}-\frac{1}{4}\right)\mathrm{.} 
\label{eq:discrete_phi}
\end{align}
Here $a, b \in \{ 0, 1, ... d-1\}$ are the measuremenmt outcomes, and $x, y = \{0,\dots,M-1\}$ are the measurement bases.   The $\mathrm{mod}(-b,d)$ term results from the definition of Bob's basis in the original CGLMP work \cite{collins_bell_2002}, and the full conditional probability is given by
\begin{equation}
P(a,b|x,y) = \frac{1}{d^3}\frac{\sin^2\left[\pi\left(a+\frac{x}{M}+ \mathrm{mod}(-b,d) + \frac{y}{M}-\frac{1}{4}  \right)\right]}{\sin^2[\frac{\pi}{d}\left(a+\frac{x}{M}+\mathrm{mod}(-b,d) + \frac{y}{M}-\frac{1}{4} \right)]}.
\label{eq: p(ab|xy)}
\end{equation}
Fig.~\ref{fig: Theoretical Simulations}c shows theoretical data from \ref{fig: Theoretical Simulations}b remapped to the form $P(a,b|x,y)$. As there are $M$ total measurement bases each with $d$ outcomes, our conditional joint probability contains a total of $(M\times d)^2$ measurement outcomes, and a subset of these are required for the CGLMP inequality to test against local hidden variables. 

In the CGLMP inequality, Alice's two bases are characterised by a contribution to the phase of her measurements given by either $\alpha = 0$ or $1/2$; the phase contributions for the two bases for Bob are given by either $\beta = 1/4$ or $-1/4$ \cite{collins_bell_2002}. Thus, the CGLMP bases are found for Alice at $x = 0 \rightarrow  x_{\alpha = 0}$ or $x = M/2 \rightarrow  x_{\alpha = 1/2}$, and for Bob at $y = M/2 \rightarrow y_{\beta = 1/4} $ or $y = 0 \rightarrow y_{\beta = -1/4}$.  These bases are represented by the white lines in Fig.~\ref{fig: Theoretical Simulations}b, where the intersections represent the joint measurements required for the CGLMP inequality.

\subsection{Experimental Results} 

Here we extend the work of ref.~\cite{serino_orchestrating_2024}, where it was shown that controllable high-dimensional entangled states of the form of equation \eqref{eq:max_entangled_state} could be generated in the temporal degree of freedom. This work used spectrally shaped pulses to generate entangled signal and idler photons in a periodically poled potassium titanyl phosphate (PPKTP) waveguide.  Analysing these results in a time-bin encoding framework, we find Alice and Bob's $\ket{j}$ states are $d$ discrete temporal modes of width $\sigma_t$ with separation $\Delta t$. The prior work focused on the generation of high-dimensional temporal-mode entanglement with controllable dimension. Here we reveal the suitability of this experiment's joint spectral intensities for tests of Bell-nonlocality using multi-outcome measurements, thus closing the binarisation loophole.

The joint spectral intensity $C(\Delta\nu_\mathrm{A}, \Delta\nu_\mathrm{A})$ for a time-bin maximally entangled state is closely related to equation \eqref{eq:theoretical_JPD_wrapped}, with the phases $\phi_\mathrm{A}$ and $\phi_\mathrm{B}$  replaced by the frequency differences $(\nu - \nu_\mathrm{A})$ and $(\nu - \nu_\mathrm{B})$ multiplied by the temporal spacing $\Delta t$ between the time-bins. The finite width of each time-bin $\sigma_t$ means the distribution in the Fourier domain is modulated by a Gaussian envelope. This is analogous to the single-slit diffraction effects noted in the Young's double-slit version \cite{li_bells_2018}.

To construct a Bell parameter $I_d$ from the joint spectral intensity data, we convert the coincidence measurements $C(\Delta\nu_\mathrm{A},\Delta\nu_\mathrm{B})$ to the conditional probabilities $P(a,b|x,y)$. 
The interference peaks in the joint spectral intensity indicate when the relative phase difference between the signal and idler modes is equal to an integer multiple of 2$\pi$, thus establishing $\phi_\mathrm{A}$ and $\phi_\mathrm{B}$.  The modulated coincidences $C(\phi_\mathrm{A},\phi_\mathrm{B})$ can then be wrapped to $C_{
\mathrm{mod}(\phi,\, 2\pi)}(\phi_\mathrm{A},\phi_\mathrm{B})$ to obtain $P_{
\mathrm{mod}(\phi,\, 2\pi)} (\phi_\mathrm{A},\phi_\mathrm{B})$.  This data is then mapped to $P(a,b|x,y)$, where a Bell parameter can be established. The wrapping of the data onto the joint probability $P_{
\mathrm{mod}(\phi,\, 2\pi)}(\phi_\mathrm{A},\phi_\mathrm{B})$ ensures that no detection events in the tails, arising from the finite width of the temporal modes, are lost and all coincidences are counted.

The joint spectral intensity is established through the implementation of coincident spectral measurements. As mentioned, this is commonly achieved in ultrafast quantum optics using dispersive fibre and a timetagger to map frequency to time. Spectral measurements are the time-frequency equivalent of simple propagation in the spatial domain, where we can use free-space propagation to convert from a near-field to a far-field~ measurement. 

\begin{figure*}
 \centering
\includegraphics{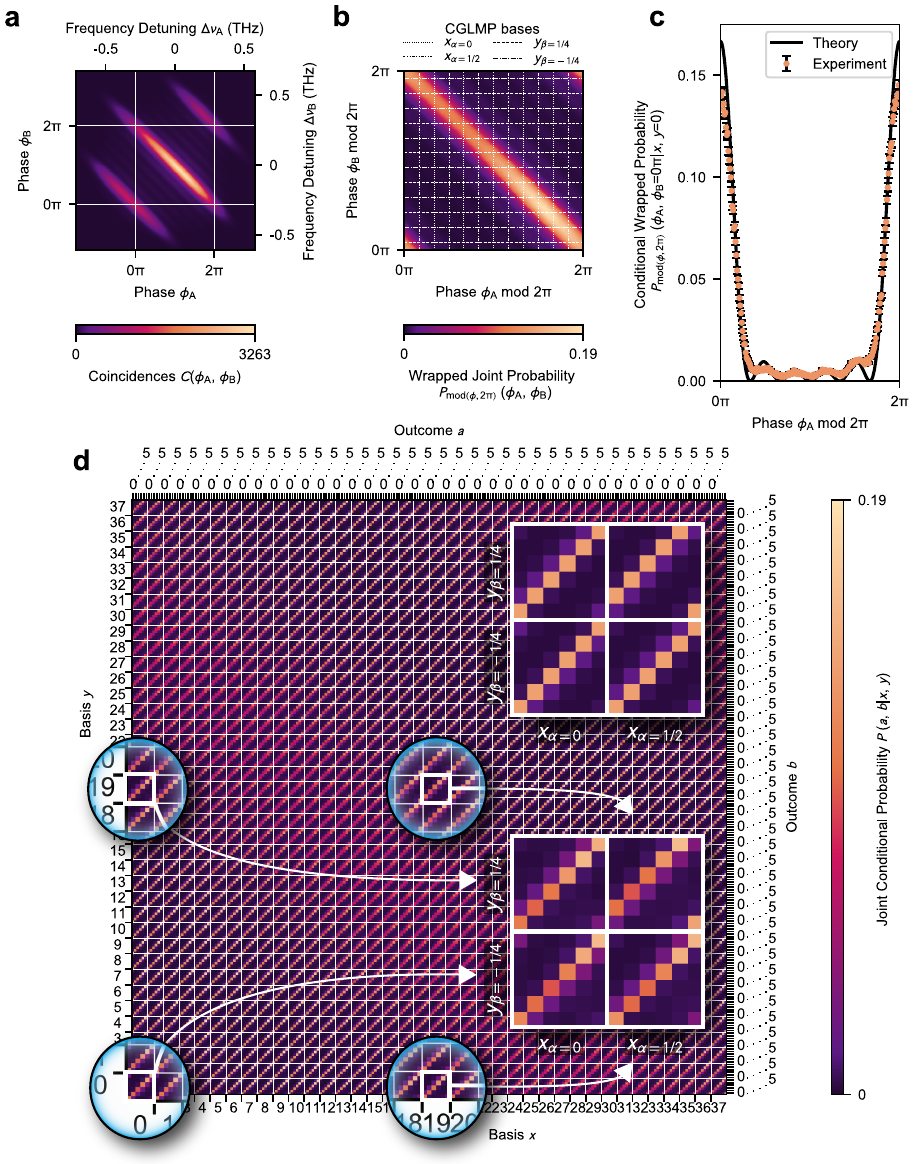}
 \caption{\textbf{Experimental results for $d = 6$. } \textbf{a}, Measured joint spectral intensity. The white lines highlight the 2$\pi$ regions, found by fitting the primary fringes. \textbf{b}, Wrapped joint probability obtained by summing the coincidence counts in the periodic $\phi$ regions of \textbf{a} and normalising them such that $\sum_{a,b=0}^{d-1} P(a,b|x,y) = 1$. The number of outcomes in a $2 \pi$ window is $N$ = 228, so that  $M$ = 38. The white lines indicate the measurements required for the CGLMP inequality. The required joint probabilities are located at the intersection of the horizontal and vertical lines.  \textbf{c}, Single row taken from \textbf{b}. The error bars were obtained by applying Poisson statistics to the JSI. \textbf{d}, The mapped probabilities from \textbf{b}, onto $P(a,b|x,y)$.  The subset of bases that are used for the CGLMP inequality are shown in the insets.  The inset in the top right is the theoretical probabilities according to Eq.~\ref{eq: p(ab|xy)}, and the inset in the bottom right is the experimentally obtained probabilities from the experiment that are used for the CGLMP inequality.}
 \label{fig: Data analysis}
\end{figure*}

\clearpage
\newpage
\onecolumngrid
\begin{center}
    \begin{figure}
 \centering
 \includegraphics{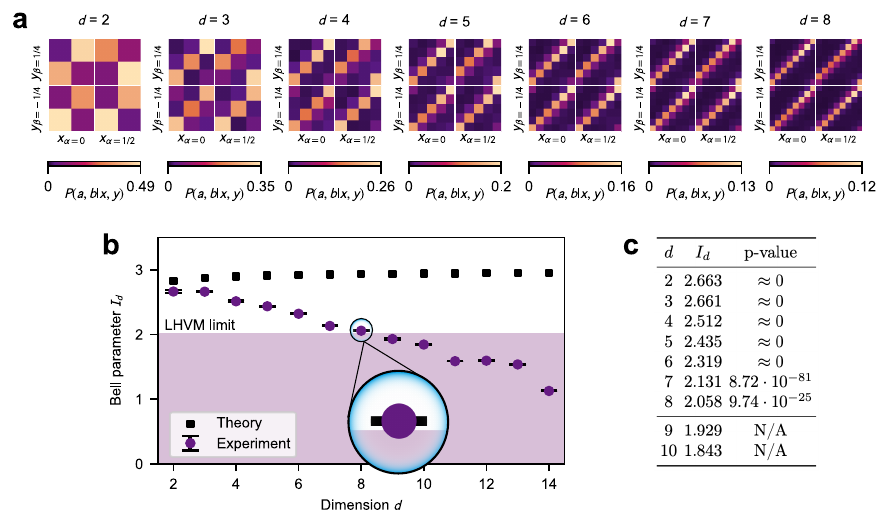}
 \caption{\textbf{Bell parameter data and metrics for $d = 2$ to 8.} \textbf{a}, Joint conditional probabilities in the CGLMP bases. \textbf{b}, Bell parameter $I_d$. For $d \leq$ 8 we find $I_d$ greater than the local hidden variable limit of 2.   The error bars represent the standard deviation of the Bell parameters, assuming the joint spectral intensities follow Poisson statistics. The theory is the Bell violation for a maximally entangled state. \textbf{c}, $I_d$ and associated $p$-values. The $p$-value for $d \leq$ 8 is $\approx 0$, thus ruling out local hidden variables with high confidence. }
 \label{fig: Bell parameter results}
 \vspace{-0.8cm}
\end{figure}
\end{center}
\twocolumngrid

%From paragraph before figure:
\noindent Just like a measurement of a momentum state in the far-field projects onto a superposition of position modes, a spectral measurement is equivalent to projecting onto time-bin superposition.

We analysed the joint spectral intensity of controllable high-dimensional entanglement, with dimension ranging from $d = 2$ to $d = 14$ \cite{serino_orchestrating_2024}, for the presence of non-local correlations.  Fig.~\ref{fig: Data analysis}a shows the joint spectral intensity for $d$~=~6.  The white lines indicate the multiple $2 \pi$ regions of $\phi_\mathrm{A}$ and $\phi_\mathrm{B}$ used to calculated the wrapped joint probability $P_{\mathrm{mod}(\phi,\, 2\pi)}(\phi_\mathrm{A},\phi_\mathrm{B})$, as shown in \ref{fig: Data analysis}b.  Fig.~\ref{fig: Data analysis}c shows a single cross-section of \ref{fig: Data analysis}b, indicating an excellent agreement between the observed data and the predictions of quantum mechanics, and \ref{fig: Data analysis}d shows the joint conditional probability $P(a,b|x,y)$ with the data used for the CGLMP inequality. Fig.~\ref{fig: Bell parameter results} shows the Bell parameter $I_d$ as calculated from the CGLMP inequality for dimensions 2 to 14.  No local hidden variable model can exceed $I_d$ = 2, thus we violate the CGLMP inequality up to $d$ = 8, all with negligible $p$-values.  The data for all dimensions that violate the CGLMP inequality are shown in the supplementary information.

There are two main factors in our experiment that reduce the Bell parameters. Firstly, the spectral filters' bandwidths are not perfectly matched to the signal and idler spectra. This results in small optical losses at the side lobes of the joint spectral intensity. Secondly, the impulse response function of the experimental system broadens the two-photon interference fringes. This impacts higher dimensions more than lower ones.

\section{Discussion}

While the CGLMP test requires only two local basis measurements, our experiment provides measurement data from many more bases. It is therefore relevant to ask whether the extra data can be used for stronger Bell inequality violations, as is known to be the case for EPR steering and entanglement \cite{Saunders_2010, Bavaresco_2018}. In Supplementary, we evidence that the CGLMP test captures most of the strength of the Bell-nonlocality signal. Specifically, we use modern convex optimisation tools \cite{Designolle_2023} to determine whether the probabilities theoretically predicted by quantum theory, after binning as in Fig.~\ref{fig: Schematic}c,  admit a local hidden variable model. By including more than two bases, we find only small improvements in the noise tolerance of the Bell-nonlocality. This justifies our focus on testing the CGLMP inequality.
 
The multi-outcome measurements obtained in our experiment close the binarisation loophole. This step is crucial to realising the noise resistance of high-dimensional states. Our approach is general and could be applied in other degrees of freedom, i.e., pixel entanglement, path, orbital angular momentum/angle, or position/momentum.  Our approach provides significant practical advantages over prior methods of establishing Bell-nonlocality, and thus, we anticipate it to be adopted for certification in quantum information protocols.  Additionally, the implications for entanglement and quantum steering remain to be explored.

This work shows that contrary to the belief that the joint spectral intensity is a phase-insensitive measurement/quantity, frequency-only measurements are sufficient to certify high-dimensional Bell-nonlocality, via the violation of the CGLMP inequality. We generate discrete high-dimensional entanglement via parametric down-conversion and measure this two-photon state after propagation through dispersive fibre. The outcomes of spectral measurements, which contain phase-sensitive superpositions of the time bins, are exactly those that are required for multi-outcome Bell tests in high dimensions. Certifying Bell-nonlocality without any measurements in the temporal domain removes one of the technological hurdles for realising high-dimensional QKD with time/frequency entanglement.

\section{Methods}

\subsection{Experimental Details}

The experiment is based on a type-II parametric down-conversion (PDC) process in an 8-mm-long periodically poled potassium titanyl phosphate (PPKTP) waveguide with a poled length of \SI{1.5}{\mm} and a poling period of \SI{117}{\micro\meter}, manufactured by AdvR Inc. The pump pulse is generated by an ultrafast Ti:Sapphire laser centred at $\lambda_\mathrm{p} = \SI{758.7}{\nano\meter}$ with a pulse duration of \SI{150}{fs} and a repetition rate of \SI{80}{\MHz}. The complex spectral amplitude of the pump is shaped by a folded-4-$f$ waveshaper \cite{monmayrant_newcomer_10} as the so-called cosine-kernel functions \cite{patera_quantum_12}, which correspond to the Fourier transform of ultrafast time-bins in the frequency domain \cite{serino_orchestrating_2024}. The PDC process generates time-frequency-entangled pairs of orthogonally polarized photons centred at $\lambda_\mathrm{0,s} = \SI{1511}{\nm}$ and $\lambda_\mathrm{0,i} = \SI{1524}{\nm}$ with a bandwidth of \SI{7}{\nm} and \SI{5}{\nm}, respectively. Note that the mismatch in signal and idler bandwidth is caused by the specific dispersion properties of the waveguide.

The remaining pump light is blocked by a band-pass filter centred at \SI{1538}{\nm} with an acceptance bandwidth of \SI{82}{\nm} (Semrock FF01-1538/82). 
The photon pair is separated by a polarizing beam splitter (PBS) and each arm is further filtered by a narrow band-pass filter with a full-width-half-maximum of \SI{7}{\nm} angle-tuned to the respective centres in order to suppress fluorescence noise. The photons are then coupled into single-mode fibres and detected by two superconducting nanowire single-photon detectors (SNSPDs) with an efficiency of 80\%. In total, we observe a Klyshko efficiency \cite{klyshko_use_80} of approximately 25\% in each arm, where most of the optical losses are caused by the fiber couplings. 

We measure the JSI using, in each arm, a time-of-flight (ToF) spectrograph \cite{avenhaus_fiber_09} consisting of a dispersive fiber with a chromatic dispersion coefficient of \SI{-418}{\ps/\nm} and a superconducting nanowire single-photon detector connected to a Swabian Instruments time-tagging unit. The ToF measurement effectively performs a Fourier transform by mapping the frequency of the photons to their arrival time.  The fibre dispersion is chosen such that effects of the finite temporal duration of the signal and idler pulses are negligible in the measured spectrum. The correlation between the arrival times of paired signal and idler photons that is, their frequency correlation, is used to reconstruct the JSI. 

% \nolinenumbers
%\printbibliography[title={References}, heading=bibintoc, resetnumbers=true]
% \linenumbers
\section{References}
\bibliography{Ref} 

\section{Acknowledgements}
K.D. and J.L. acknowledge financial support from EPSRC projects ID EP/Z533178/1 and EP/Z533166/1.  
N.A and A.T are supported by the Swedish Research Council under Contract No. 2023-03498 and the Knut and Alice Wallenberg Foundation through the Wallenberg Center for Quantum Technology (WACQT).
L.S., A.B., B.B., and C.S. acknowledge funding from the European Union's Horizon Europe research and innovation programme under Grant Agreement No. 101070700 (MIRAQLS).

\end{document}

\title{Supplementary Information: Observing High-Dimensional Bell Inequality Violations using Multi-Outcome Spectral Measurements}

\author{Kiki Dekkers}
\affiliation{School of Engineering and Physical Sciences, Heriot-Watt University, Edinburgh, EH14 4AS, UK}

\author{Laura Serino}
\affiliation{Paderborn University, Integrated Quantum Optics, Institute for Photonic Quantum Systems (PhoQS), Paderborn, Germany}

\author{Nicola D'Alessandro}
\affiliation{Physics Department and NanoLund, Lund University, Box 118, 22100 Lund, Sweden\\ $^*$j.leach@hw.ac.uk}

\author{Abhinandan Bhattacharjee}
\affiliation{Paderborn University, Integrated Quantum Optics, Institute for Photonic Quantum Systems (PhoQS), Paderborn, Germany}

\author{Benjamin Brecht}
\affiliation{Paderborn University, Integrated Quantum Optics, Institute for Photonic Quantum Systems (PhoQS), Paderborn, Germany}

\author{Armin Tavakoli}
\affiliation{Physics Department and NanoLund, Lund University, Box 118, 22100 Lund, Sweden\\ $^*$j.leach@hw.ac.uk}

\author{Christine Silberhorn}
\affiliation{Paderborn University, Integrated Quantum Optics, Institute for Photonic Quantum Systems (PhoQS), Paderborn, Germany}

\author{Jonathan Leach}
%\email{j.leach@hw.ac.uk}
\affiliation{School of Engineering and Physical Sciences, Heriot-Watt University, Edinburgh, EH14 4AS, UK}

\maketitle

\section{CGLMP inequality}
The CGLMP Bell inequality \cite{collins_bell_2002} is a facet of the local polytope in the scenario in which Alice and Bob have binary inputs $x,y\in\{1,2\}$ and respective outcomes $a,b\in\{0,\ldots,d-1\}$, for some integer $d\geq 2$. For any local hidden variable theory, it holds that
\begin{align}\nonumber
    I_d=\sum_{k=0}^{\lfloor d/2\rfloor -1} \left(1-\frac{2k}{d-1}\right)& \big[P(a_1=b_1+k)+P(b_1=a_2+k+1)+P(a_2=b_2+k)+P(b_2=a_1+k)\\
    &-P(a_1=b_1-k-1)-P(b_1=a_2-k)-P(a_2=b_2-k-1)-P(b_2=a_1-k-1)\big]\leq 2,
    \label{eq: supl. CGLMP inequality}
\end{align}
where we have defined
\begin{equation}
    P(a_x=b_y+z)=\sum_{a,b=0} ^{d-1}P(a,b|x,y)\delta_{a_x, \ b_y+z \ \mathrm{mod} \ d},
\end{equation}
and
\begin{equation}
    P(b_y=a_x+z)=\sum_{a,b=0} ^{d-1} P(a,b|x,y)\delta_{a_x+z  \ \mathrm{mod} \ d,\ b_y}.
\end{equation}
For the special case of $d=2$ the CGLMP inequality reduces to the CHSH inequality.\\
\\
To violate the CGLMP inequality, Alice and Bob can share the maximally entangled state
\begin{equation}
\ket{\psi} = \frac{1}{\sqrt{d}} \sum_{j=0}^{d-1} \ket{j}_A \ket{j}_B, 
\label{eq:max_entangled_state}
\end{equation}
The best-known measurement selections correspond to the following bases
\begin{align}
&\ket{K_{k|x}} = \frac{1}{\sqrt{d}} \sum_{j=0}^{d-1} \exp\left( i \frac{2\pi}{d} j(k + \alpha_x) \right) \ket{j}_A~\textrm{and} \\
& \ket{L_{l|y}} = \frac{1}{\sqrt{d}} \sum_{j=0}^{d-1} \exp\left( i \frac{2\pi}{d} j(-l + \beta_y) \right) \ket{j}_B, 
\label{eq:Bell_basis_A}
\end{align}
where $\alpha_1 = 0, \alpha_2 = 1/2, \beta_1 = 1/4$ and $\beta_2 = -1/4$. These are the eigenvectors from the Fourier and inverse Fourier bases as defined in the original CGLMP inequality \cite{collins_bell_2002}.  This gives rise to a joint probability of 
\begin{align}
P(k, l| x, y) = \frac{1}{d^3} \frac{\sin^2 \left[\pi(k +\alpha_x) + \pi(-l + \beta_y) \right]}{\sin^2 \left[ \frac{\pi}{d}(k +\alpha_x) + \frac{\pi}{d}(-l + \beta_y)  \right]}.
\end{align}
Here we have grouped Alice's $(k +\alpha_x)$ and Bob's $(-l + \beta_y)$ terms together. The $k$s and $l$s are the measurement outcomes, given a choice of basis $x = 1,2$ for Alice and $y = 1,2$ for Bob.

In our work, both Alice and Bob both perform measurements in the Fourier basis, with measurements corresponding to 
\begin{align}
\ket{\phi} = \frac{1}{\sqrt{d}} \sum_{j=0}^{d-1} \exp\left(- i j \phi \right) \ket{j}.
\end{align}
Note that in the original CGLMP work, the Fourier transform is defined without the negative sign. The complex overlap is then 
\begin{align}
\bra{\phi}_A\bra{\phi}_B\ket{\psi} & = \frac{1}{\sqrt{d}} \sum_{j=0}^{d-1} \exp\left(i j \phi_A \right) \bra{j}_A \frac{1}{\sqrt{d}} \sum_{j=0}^{d-1} \exp\left(i j \phi_B \right) \bra{j}_B \frac{1}{\sqrt{d}} \sum_{j=0}^{d-1} \ket{j}_A \ket{j}_B\\
& = \frac{1}{\sqrt{d^{3}}}\sum_{j=0}^{d-1} \exp\left(i j (\phi_A +\phi_B) \right).
\end{align}
The joint probability parameterised by $\phi_A$ and $\phi_B$ that we measure is given by
\begin{align}
P(\phi_A, \phi_B) & = |\bra{\phi}_A\bra{\phi}_B\ket{\psi}|^2\\
& = \frac{1}{d^{3}}\left|\sum_{j=0}^{d-1} \exp\left(i j (\phi_A +\phi_B) \right)\right|^2\\
& = \frac{1}{d^3} \frac{\sin^2 \left[d(\phi_\mathrm{A}+\phi_\mathrm{B})/ 2 \right]}{\sin^2 \left[ (\phi_\mathrm{A}+\phi_\mathrm{B}) / 2 \right]}\\
& =\frac{1}{d^3} \frac{\sin^2 \left[\frac{d}{2}\phi_\mathrm{A}+\frac{d}{2}\phi_\mathrm{B} \right]}{\sin^2 \left[ \frac{1}{2}\phi_\mathrm{A}+\frac{1}{2}\phi_\mathrm{B} \right]}.
\end{align}
In practice, the finite resolution of our measurement system means that we record discrete phases.  If, in any $2\pi$ window, an experiment records $N$ outcomes, these can be identified as the $d$ outcomes from $M$ measurement bases, i.e.~$N = M\times d$. 

The experimentally measured coincidences at the phases $\phi_A$ and $\phi_B$ can be mapped to the conditional probabilities required for the CGLMP inequality via
\begin{equation}
\phi_{A} \rightarrow \frac{2\pi}{d}\left(a + \frac{x}{M}\right) {\rm~and}~\phi_{B} \rightarrow \frac{2\pi}{d}\left(\mathrm{mod}(-b,d) + \frac{y}{M}-\frac{1}{4}\right)\mathrm{,} 
\label{eq:discrete_phi}
\end{equation}
where $a = \{ 0, 1, ... d-1\}$ is the measurement outcome index for Alice, and $x = \{0, 1, ..., M-1\}$ is the measurement bases index for Alice, and Bob's measurement is indexed by $b$ and $y$. This gives rise to the conditional probability 
\begin{equation}
P(a,b|x,y) = \frac{1}{d^3}\frac{\sin^2\left[\pi\left((a+\frac{x}{M})+ (\mathrm{mod}(-b,d) + \frac{y}{M}-\frac{1}{4})  \right)\right]}{\sin^2[\frac{\pi}{d}\left((a+\frac{x}{M})+ (\mathrm{mod}(-b,d) + \frac{y}{M}-\frac{1}{4}) \right)]},
\label{eq: p(ab|xy)}
\end{equation}
where the measurement vectors are
\begin{align}
&\ket{A_{a|x}} = \frac{1}{\sqrt{d}} \sum_{j=0}^{d-1} \exp\left( - i \frac{2\pi}{d} j \left(a + \frac{x}{M} \right) \right) \ket{j}_A~\textrm{and}\\
& \ket{B_{b|y}} = \frac{1}{\sqrt{d}} \sum_{j=0}^{d-1} \exp\left( - i \frac{2\pi}{d} j \left(\mathrm{mod}(-b,d) + \frac{y}{M}-\frac{1}{4}\right) \right) \ket{j}_B. 
\label{eq:Bell_basis_A}
\end{align}
We see that the outcomes $k$ and $l$ from the CGLMP inequality are the outcomes $a$ and $b$ in our work, and the additional contributions to the phases $\alpha_x$ and $\beta_y$ required for the basis choices are the terms $x/M$ and $y/M-1/4$. Note that in the CGLMP work $x,y$ are the binary inputs $x,y\in\{1,2\}$, whereas in our work we measure in multiple bases corresponding to     $x,y\in\{0,1,...,M-1\}$.   Whether the Fourier transform is defined with or without the negative sign results in the same joint conditional probability, equation \ref{eq: p(ab|xy)}.

\section{Data analysis}
\subsection{Mapping the joint spectral intensity $C(\Delta\nu_\mathrm{A}, \Delta\nu_\mathrm{B})$ to the wrapped joint probability $P_{\mathrm{mod}(\phi,\, 2\pi)} (\phi_\mathrm{A}, \phi_\mathrm{B})$}\label{supl inf: wrapping JSI}
\begin{figure}
    \centering
    \includegraphics{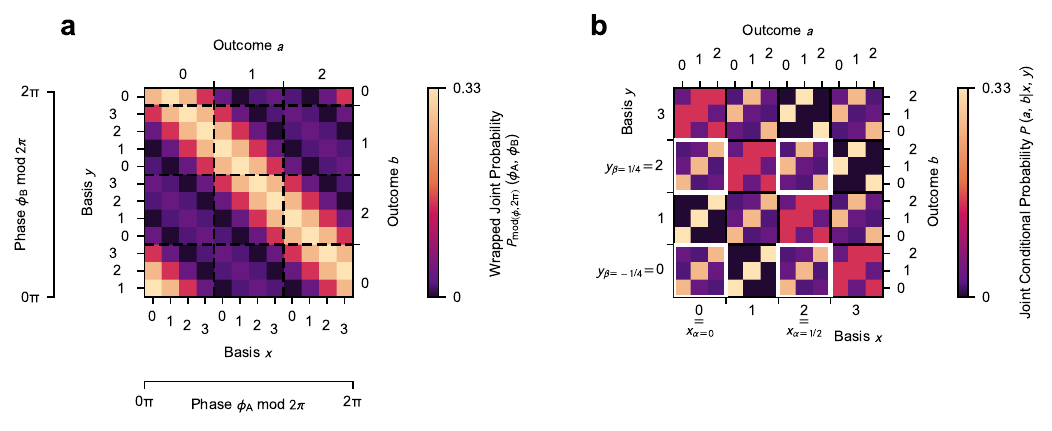}
    \caption{\textbf{Visualisation of relationship between, $\phi_\mathrm{A}$ ($\phi_\mathrm{B}$), and, $a$ ($b$) and $x$ ($y$) for $d=3$ and $M=5$.} \textbf{a}, Theoretical wrapped joint probability $P_{ \mathrm{mod(}\phi, \ 2\uppi)} (\phi_\mathrm{A}, \phi_\mathrm{B})$. \textbf{b}, Joint probability distribution sorted by measurement bases.}
    \label{fig: Relationship_phi_a_b_x_y}
\end{figure}

Whenever a function has a finite extent, its Fourier transform is modulated by an envelope. For example, single-slit diffraction modulates the double-slit interference pattern in Young's double slit experiment. Similarly, in our system, the joint spectral intensity is modulated by a Gaussian envelope due to the finite widths $\sigma_t$ of our Gaussian time bins. 
Now consider the case where we are interested in the joint probability distribution $P(\phi_\mathrm{A}, \phi_\mathrm{B})$ in the Fourier domain of our entangled state. The amplitude of the joint probability distribution $P(\phi_\mathrm{A}, \phi_\mathrm{B})$ decreases as we move away from the distribution centre. Therefore, the modulated distribution over one phase period does not accurately represent the sought after unmodulated $P(\phi_\mathrm{A}, \phi_\mathrm{B})$. 
Li and Zhao \cite{li_bells_2018} showed that when only using one phase period of the modulated $P(\phi_\mathrm{A}, \phi_\mathrm{B})$ to extract the necessary the CGLMP measurements, the maximal theoretically possible Bell parameter decreases - in some cases to even below the LHVM limit. They propose mitigating this effect by carefully selecting experimental parameters to suppress the envelope modulation of $P(\phi_\mathrm{A}, \phi_\mathrm{B})$. However, we recognise that the unmodulated $P(\phi_\mathrm{A}, \phi_\mathrm{B})$ distribution can be recovered by mapping the modulated $P(\phi_\mathrm{A}, \phi_\mathrm{B})$ onto the wrapped joint probability $P_{\mathrm{mod}(\phi,\, 2\pi)}(\phi_\mathrm{A}, \phi_\mathrm{B})$. This ensures that no outcomes are discarded. \\
\\
This mapping of the joint spectral intensity $C(\Delta\nu_\mathrm{A}, \Delta\nu_\mathrm{B})$ onto $P_{\mathrm{mod}(\phi,\, 2\pi)}(\phi_\mathrm{A}, \phi_\mathrm{B})$ is achieved as follows. The boundaries of the 2$\uppi$ phase periods are found by fitting three diagonal Gaussians to the principal interference fringes in the $C(\Delta\nu_\mathrm{A}, \Delta\nu_\mathrm{B})$ data. Their separation was restricted to an integer multiple of $d$ to ensure an integer number of measurement bases $M$. We infer $C(\phi_\mathrm{A}, \ \phi_\mathrm{B})$ from the best fit of the principal fringes separation parameter. Now that the cyclic 2$\uppi$ square regions have been identified, we stack these regions, i.e. sum them, to obtain the wrapped $C_{\mathrm{mod}(\phi,\, 2\pi)}(\phi_\mathrm{A}, \phi_\mathrm{B})$. This coincidence distribution is normalised to $P_{\mathrm{mod}(\phi,\, 2\pi)}(\phi_\mathrm{A}, \phi_\mathrm{B})$ such that for each combination of measurement basis $x$ and $y$, the total probability for all measurement outcome combinations $a$ and $b$ sums to 1, i.e. $\sum_{a,b=0}^{d-1} P(a,b|x,y) = 1$. The relationship between $\phi_\mathrm{A}$ ($\phi_\mathrm{B}$), and $a$ ($b$) and $x$ ($y$) is given by equation 3 (4). For a visualisation of these relationships, see Fig.~\ref{fig: Relationship_phi_a_b_x_y}.

% \subsection{Quantum Contrast}
% ...

\subsection{Poisson statistics}
Assuming the joint spectral intensity coincidence detection events $C$ follow Poisson statistics, its standard deviation $\sigma$ is given by $\sigma = \sqrt{C}$. \\
\\
The standard deviation on the Bell parameters $I_d$ in Fig. 5b in the main paper was determined as follows for each dimension $d$. A Poisson distribution was created where the expected number of events occurring was set by the wrapped coincidence distribution $C_{\mathrm{mod}(\phi,\, 2\pi)}(\phi_\mathrm{A}, \phi_\mathrm{B})$. The Poisson distribution was then normalised to $P_{\mathrm{mod}(\phi,\, 2\pi)}(\phi_\mathrm{A}, \phi_\mathrm{B})$ and its Bell parameter was extracted. This was repeated 50 times. The standard deviation on each Bell parameter $I_d$ of the experimental data was calculated based on the set of the Bell parameters of its respective 50 Poisson distributions. It was checked that 50 repeats were sufficient as increasing the number of Poisson distribution repeats did not impact the magnitude of the standard deviation significantly.

\subsection{Statistical significance}
The estimate of the Bell parameter indicates a violation of the CGLMP inequality in dimensions up to eight. Here, we discuss the statistical significance of these violations and show that they all come with vanishingly small $p$-values. To quantify the statistical significance, we need to evaluate the probability that a local hidden variable model (LHVM) could reproduce the measured finite-statistics values of the CGLMP parameter. The probability of statistical fluctuations in the LHVM explaining the measured violations is called the $p$-value. Thus, a small $p$-value corresponds to a high confidence in observing Bell-nonlocality.

To estimate the $p$-values for the violations obtained in the experiment, we use the method described in \cite{Elkouss_2016}. We begin considering a general Bell inequality in the form
\begin{equation}
    \mathcal{B}=\sum_{x,y,a,b}s_{ab}^{xy}P(ab|xy)\leq \beta
\end{equation}
and estimate the probability of selecting settings $(x,y)$ from the photon coincidences counts $N_{a,b,x,y}$ detected when obtaining outcome $(a,b)$
\begin{equation}
    P(x,y)=\dfrac{\sum\limits_{a,b}N_{a,b,x,y}}{\sum\limits_{a,b,x,y}N_{a,b,x,y}}.
\end{equation}
The inequality can then be rewritten as
\begin{equation}
    \mathcal{B}=\sum_{x,y,a,b}s_{abxy}P(x,y)P(ab|xy).
\end{equation}
In this form, $s_{abxy}$ can be viewed as the weighted score that the observers obtain when they answer $a$ and $b$ to the question $x$ and $y$. Any round of the Bell game then corresponds to evaluating a cost function whose value is between $s_{max}=\max\limits_{a,b,x,y} s_{abxy}$ and $s_{min}=\min\limits_{a,b,x,y} s_{abxy}$. The total score that the observers obtain after $n$ rounds of the experiment will then be
\begin{equation}
    c=\sum_{i=1}^ns_{a_ib_ix_iy_i}
\end{equation}

This form is favourable because the data can now be interpreted in terms of an $n$-round  game in which a certain score is awarded in every round. Since $s_{abxy}=s_{ab}^{xy}/P(x,y)$, this score is influenced by the settings probability distribution.
%The key idea of this method, is then to use these total scores to compute the probability that a LHVM could produce results at least as large as those observed, thus yielding a P-value that quantifies our confidence in the nonlocal nature of the data.
In order to bound the $p$-value associated with the measured score, we use McDiarmid’s inequality \cite{McDiarmid1989SurveysIC}
\begin{equation}\label{Eq:McDir}
    \text{$p$-value}\leq \left( \left( \dfrac{s_\text{max}-\beta}{s_\text{max}-c/n}\right)^{\dfrac{s_\text{max}-c/n}{s_\text{max}-s_\text{min}}} \left( \dfrac{\beta-s_\text{min}}{c/n-s_\text{min}}\right)^{\dfrac{c/n-s_\text{min}}{s_\text{max}-s_\text{min}}} \right)^n.
\end{equation}
We apply this  bound to the experiment's data. Due to the high number of photon coincidences obtained with our setup and the comparatively large estimated inequality violations, all $p$-values up to dimension 6 are all below $10^{-200}$. For dimensions $7$ and $8$, the smaller estimated inequality violations lead to larger $p$-values but these are still vanishingly small. Specifically, they are $8.7\cdot 10^{-81}$ and $9.7\cdot 10^{-25}$ for $d=7$ and $d=8$ respectively.

\subsection{More Bell-nonlocality from the same data?}
The experiment provides measurement statistics in many local bases. It is therefore natural to ask whether it is reasonable to focus on the CGLMP inequality given that this test only uses two bases per party. 

In general, the Bell-nonlocality of any given bipartite probability distribution can be assessed by means of convex programming techniques. In principle, this offers an exact solution. If the distribution is found to be Bell-nonlocal, one can compute a suitable distance measure to the set of correlations realisable with LHVM. One can also use the duality theory of convex programming to extract a Bell inequality that is violated by the target distribution \cite{Tavakoli_2024}. However, all this rapidly becomes expensive as the number of inputs and outputs grows, which is the case in our experiment.  

To overcome this obstacle, we have used efficient approximation methods for determining whether the distribution is Bell-nonlocal, and if so extracting the relevant Bell inequality. Our approach is based on the Frank-Wolfe algorithm \cite{FW_1, FW_2} and closely follows the method developed in \cite{Designolle_2023}. 

As shown in Tab. \ref{tab:p_val}, by constructing Bell inequalities tailored to our measured data, we obtain sharper bounds in scenarios with two measurement settings per party, improving the statistical significance beyond the CGLMP result. Furthermore, when extending the method to more than two measurements settings, we observe no practical benefit. In fact, the noise tolerance remains practically the same, only decreasing from $v_2=0.7423$ to $v_{30}=0.7411$ when moving to $30$ settings in dimension $2$. A similar behaviour was observed in higher dimensions. For example, in dimension $8$ the critical resistance to noise remains unchanged, up to numerical resolution, when increasing the settings from $2$~to~$9$. Additionally, the statistical penality of adding more settings to the test overshadows those minor gains, ultimately increasing the $p$-value in all cases.

A natural question could be whether the lack of improvements stems from using the theoretical probability distribution. To investigate this, we tested Bell inequalities obtained when using as target probability distribution the no-singaling point closest to the experimental data. Since the results were nearly identical to those obtained with the theoretical distribution, we do not elaborate further on this analysis.

\begin{table}[h!]
\centering
\begin{tabular}{|c|c|c|c|c|}
\hline
   $d$ & $ v_\text{CGLMP} $  & $ v_\text{opt} $ & $\text{$p$-value}_{\text{CGLMP}}$ & $\text{$p$-value}_{\text{opt}}$ \\ \hline
$2$ & 0.7499 & 0.7482 & $ \approx 0$ &  $\approx 0$ \\ \hline
$3$ & 0.7489 & 0.7423 & $ \approx 0$ &  $\approx 0$ \\ \hline
$4$ & 0.7963 & 0.7800 & $ \approx 0$ &  $\approx 0$ \\ \hline
$5$ & 0.8239 & 0.8129 & $ \approx 0$ &  $\approx 0$ \\ \hline
$6$ & 0.8616 & 0.8615 & $ \approx 0$ &  $\approx 0$ \\ \hline
$7$ & 0.9346 & 0.9331 & $8.72\cdot 10^{-81}$ & $ 3.04 \cdot 10^{-83}$ \\ \hline
$8$ & 0.9692 & 0.9640 & $9.74\cdot 10^{-25}$ & $1.35\cdot 10^{-34}$ \\ \hline
\end{tabular}
\caption{Critical visibilities ($ v_\text{CGLMP}, \, v_\text{opt}$) and corresponding $p$-values ($p\text{-value}_\text{CGLMP}$, $p\text{-value}_\text{opt}$), obtained when testing the CGLMP ineqality and the optimal two settings inequality found numerically for dimensions $d=\{2,\dots, 8\}$.}
\label{tab:p_val}
\end{table}

\section{Experimental data}

\begin{figure}
    \centering
    \includegraphics{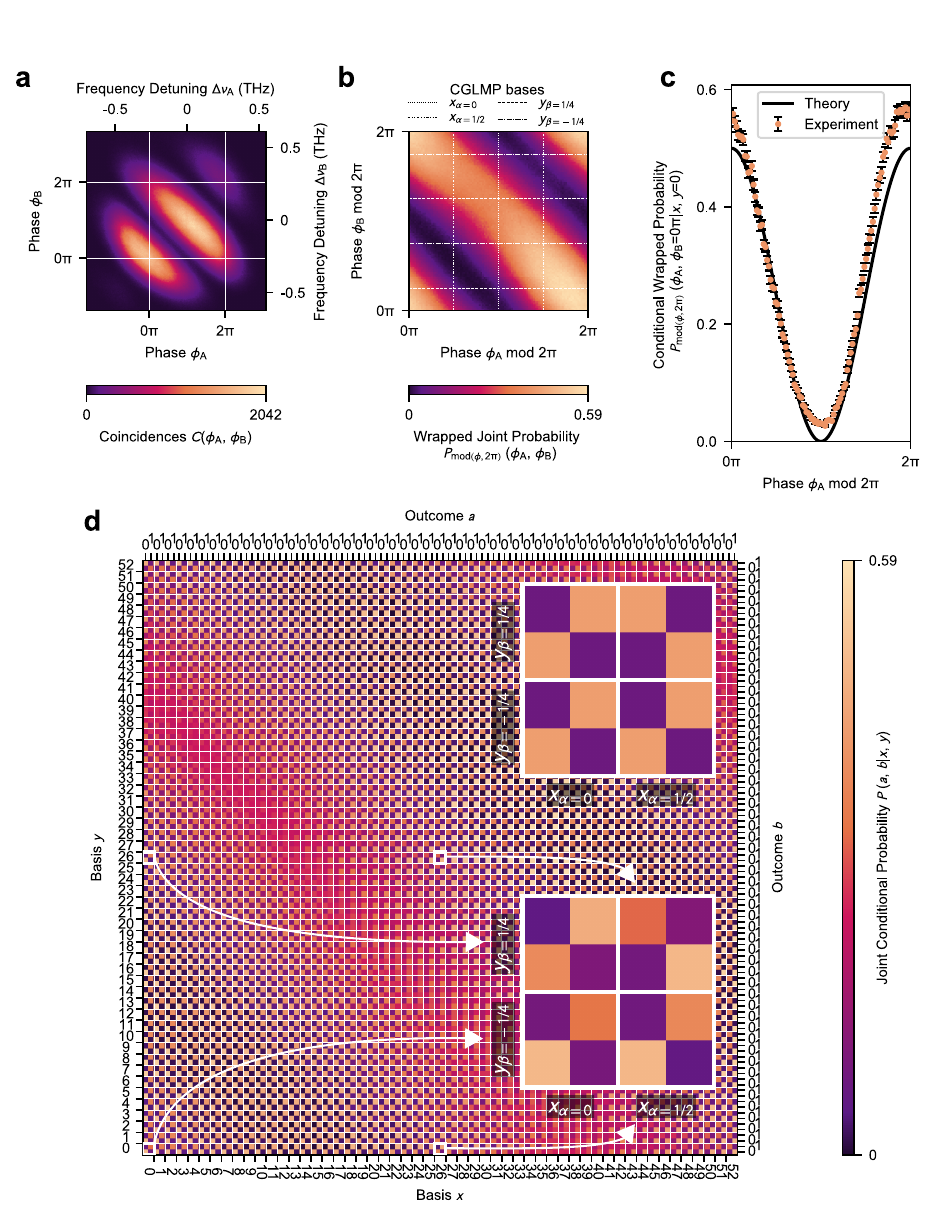}
    \caption{Data for $d =2$.}
    \label{fig:enter-label}
\end{figure}

\begin{figure}
    \centering
    \includegraphics{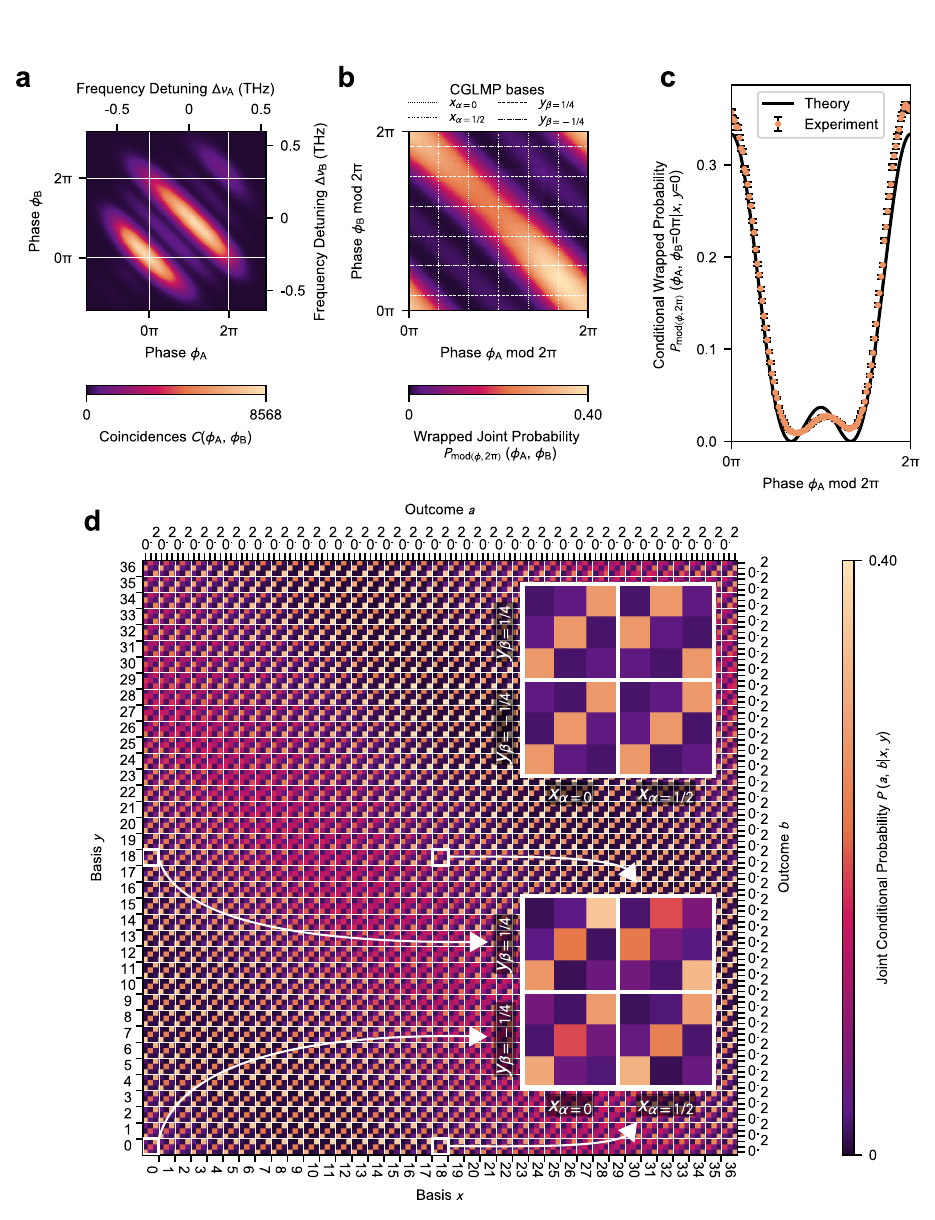}
    \caption{Data for $d =3$.}
    \label{fig:enter-label}
\end{figure}

\begin{figure}
    \centering
    \includegraphics{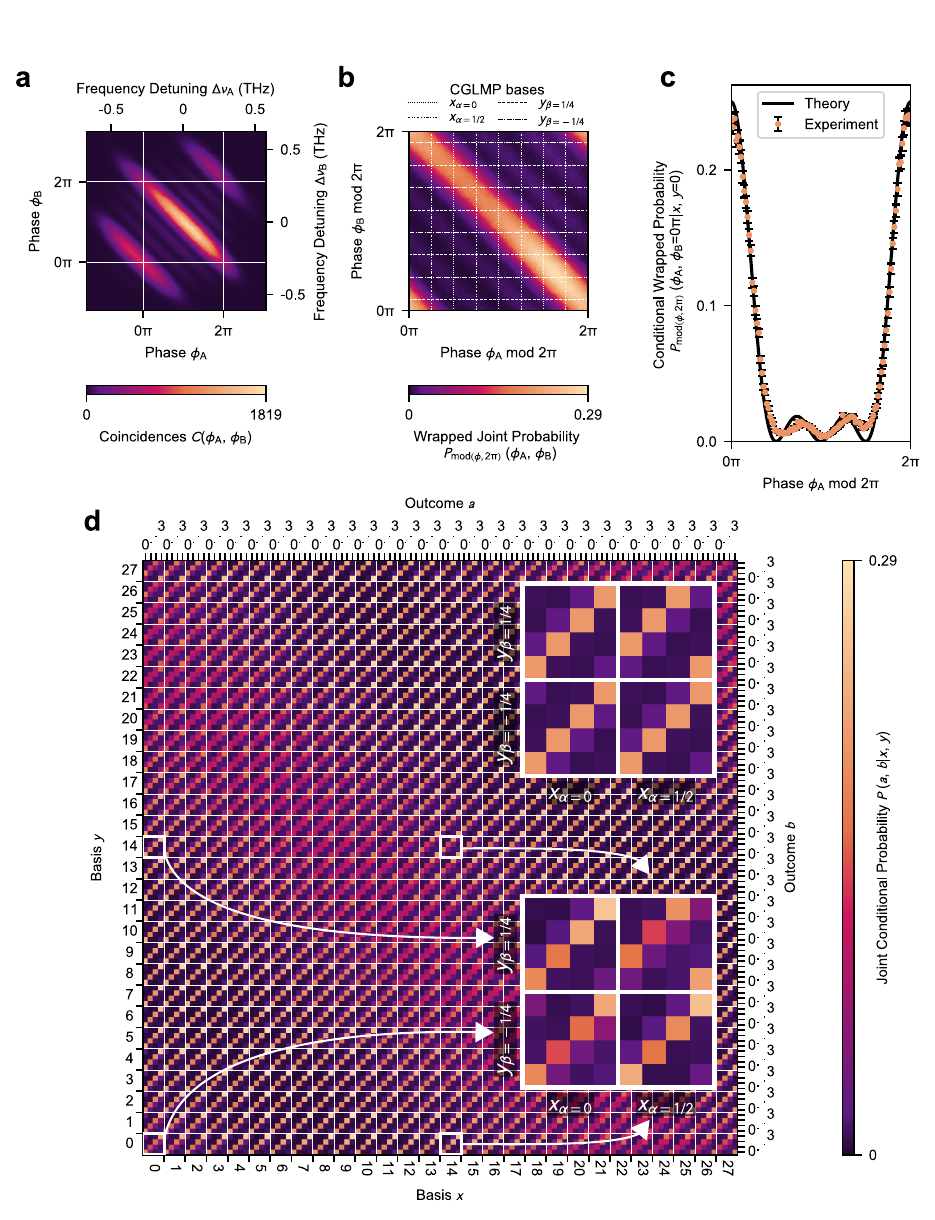}
    \caption{Data for $d =4$.}
    \label{fig:enter-label}
\end{figure}

\begin{figure}
    \centering
    \includegraphics{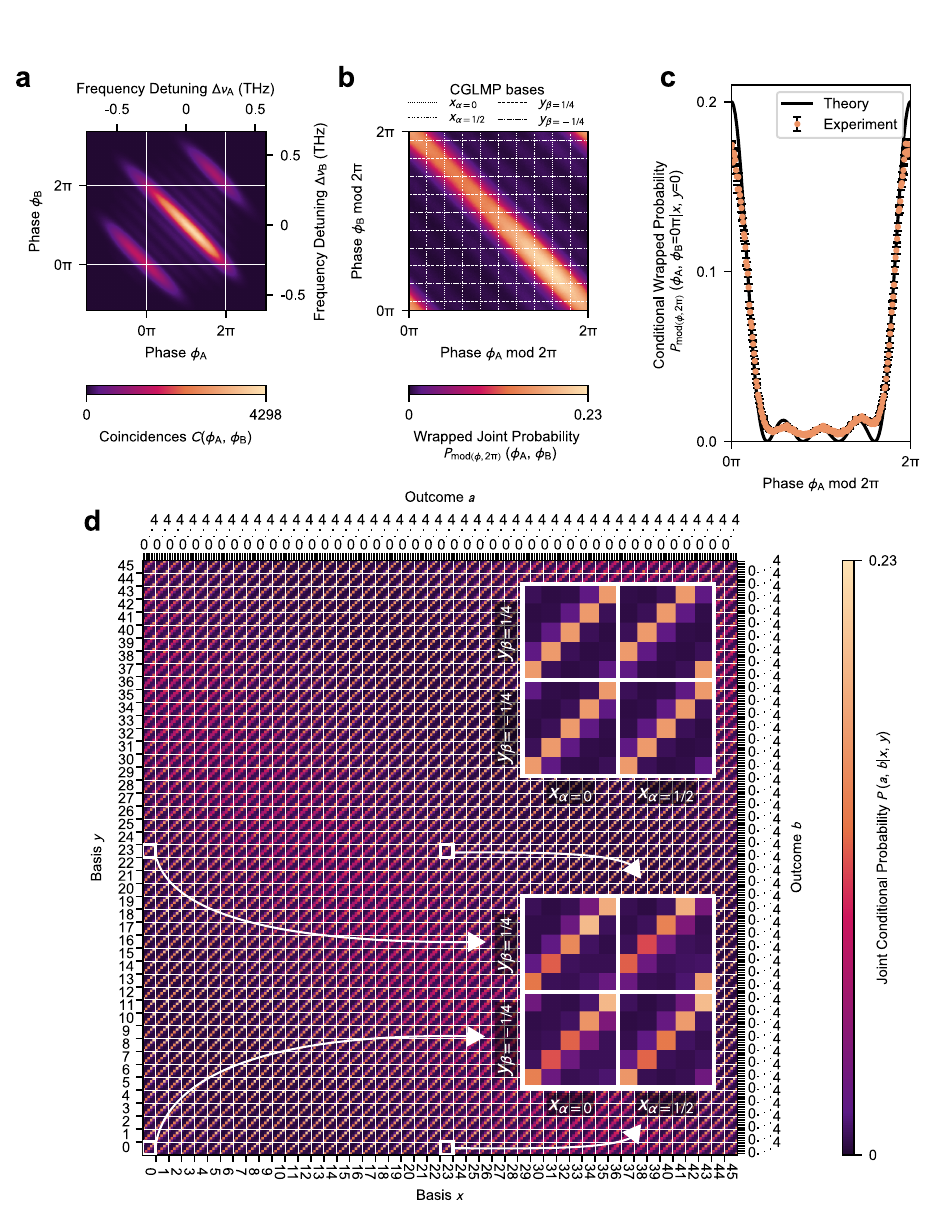}
    \caption{Data for $d =5$.}
    \label{fig:enter-label}
\end{figure}

\begin{figure}
    \centering
    \includegraphics{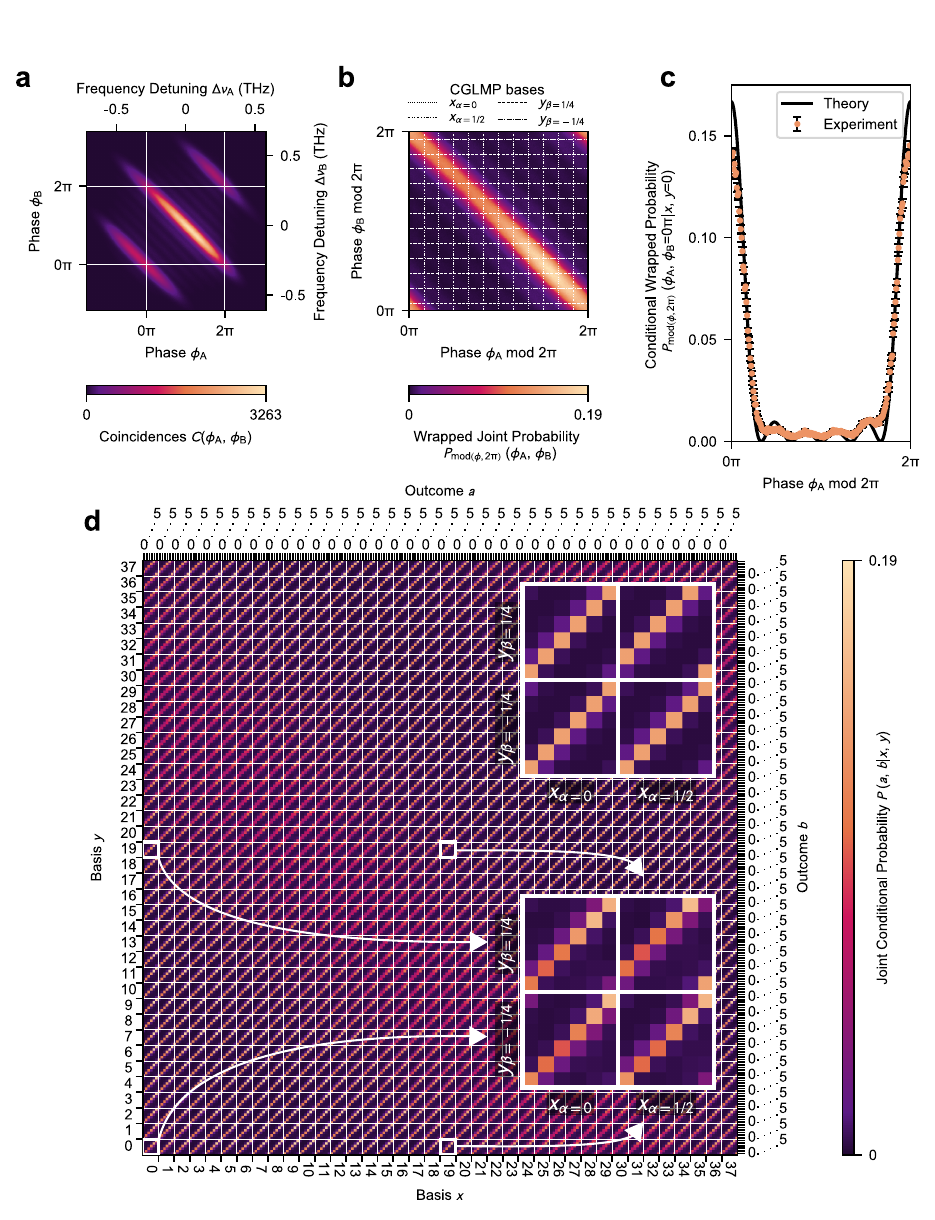}
    \caption{Data for $d =6$.}
    \label{fig:enter-label}
\end{figure}

\begin{figure}
    \centering
    \includegraphics{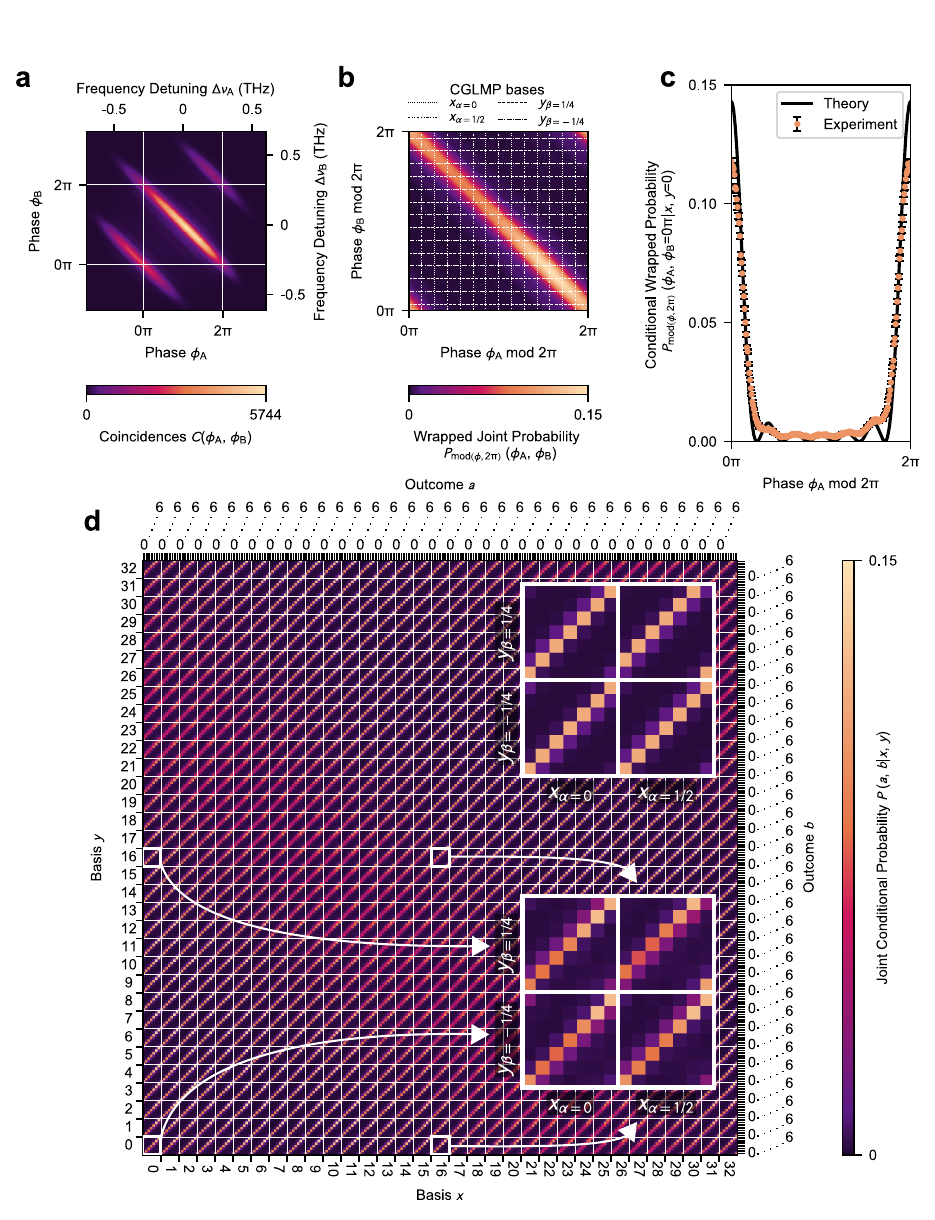}
    \caption{Data for $d =7$.}
    \label{fig:enter-label}
\end{figure}

\begin{figure}
    \centering
    \includegraphics{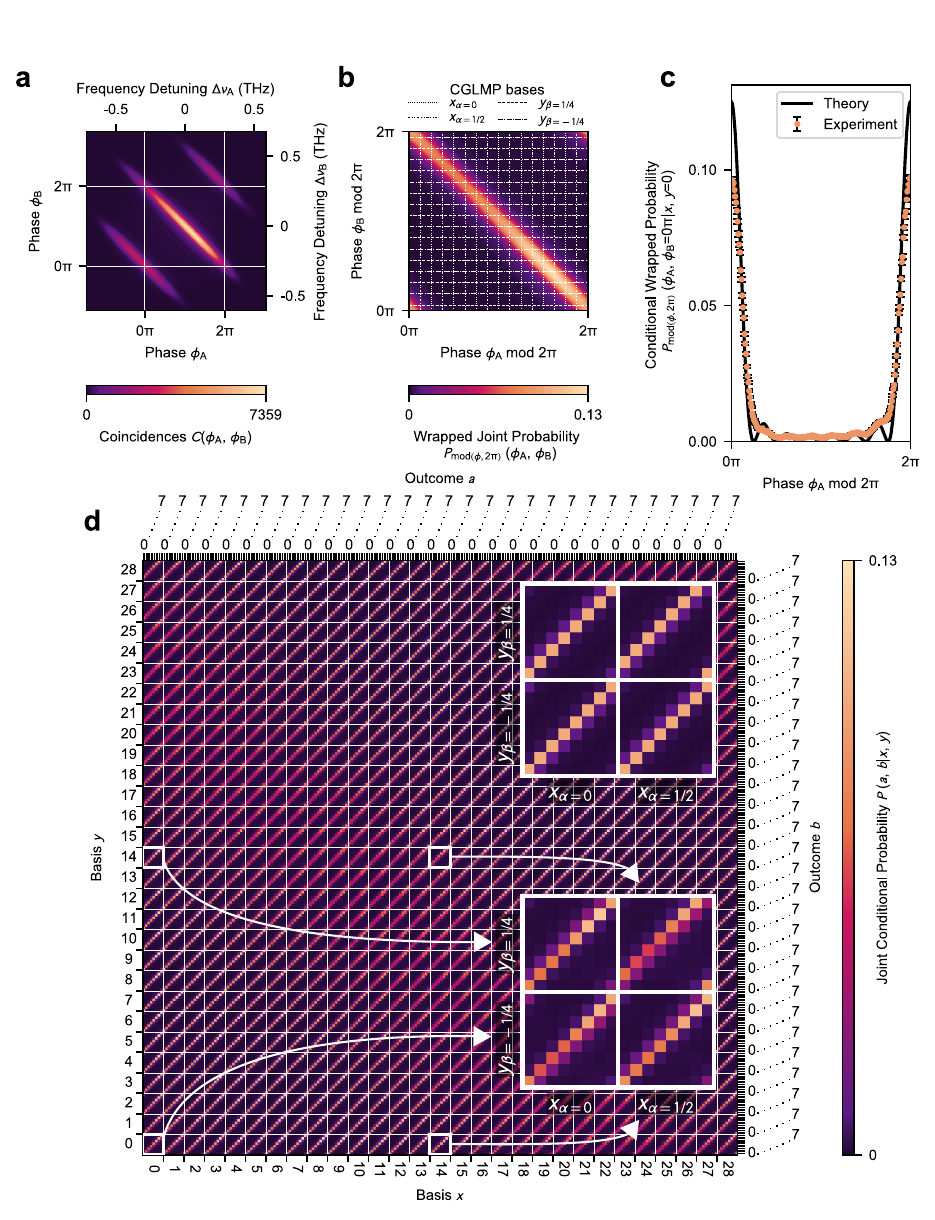}
    \caption{Data for $d =8$.}
    \label{fig:enter-label}
\end{figure}

\newpage

\section{References}
\bibliography{Ref2}